\documentclass[aip,jap,reprint,superscriptaddress,floatfix]{revtex4-1}

\usepackage{graphicx,amsmath,amssymb,multirow,rotate,color,subfigure}
\usepackage{ae,aecompl}
\usepackage{hyperref}
\usepackage{txfonts}

\usepackage[utf8]{inputenc}  
\usepackage[T1]{fontenc}       

\usepackage{color}
\definecolor{red}{rgb}{1,0,0}
\definecolor{green}{rgb}{0,1,0}
\definecolor{blue}{rgb}{0,0,1}

\graphicspath{{figsAcciolySR/}}

\makeatletter
\hypersetup{
		pdftitle={Role of STTs on Sync and Resonance Phenomena in Stochastic Magnetic Oscillators}, 
		pdfauthor={A. Accioly et al.},
		colorlinks=true,       		
    	linkcolor=blue,          	
    	citecolor=blue,        		
    	filecolor=magenta,      		
		urlcolor=blue,
		bookmarksdepth=4
}
\makeatother

\begin{document}

\title{Role of spin-transfer torques on synchronization and resonance phenomena in stochastic magnetic oscillators}

\author{Artur~\surname{Accioly}}
\affiliation{Instituto de F\'\i sica, 
             Universidade Federal do Rio Grande do Sul, 91501-970 Porto Alegre, Brazil}
\affiliation{Centre de Nanosciences et de Nanotechnologies, CNRS, Univ. Paris-Sud, Universit{\'e} Paris-Saclay, 91405 Orsay, France}    
\author{Nicolas~\surname{Locatelli}}
\affiliation{Centre de Nanosciences et de Nanotechnologies, CNRS, Univ. Paris-Sud, Universit{\'e} Paris-Saclay, 91405 Orsay, France}    
\author{Alice~\surname{Mizrahi}}
\affiliation{Centre de Nanosciences et de Nanotechnologies, CNRS, Univ. Paris-Sud, Universit{\'e} Paris-Saclay, 91405 Orsay, France}
\affiliation{Unit{\'e} Mixte de Physique CNRS, Thales, Univ. Paris-Sud, Universit{\'e} Paris-Saclay, F91767 Palaiseau, France}             
\author{Damien~\surname{Querlioz}}
\affiliation{Centre de Nanosciences et de Nanotechnologies, CNRS, Univ. Paris-Sud, Universit{\'e} Paris-Saclay, 91405 Orsay, France}    
\author{Luis G.~\surname{Pereira}}
\affiliation{Instituto de F\'\i sica, 
             Universidade Federal do Rio Grande do Sul, 91501-970 Porto Alegre, Brazil}             
\author{Julie~\surname{Grollier}}
\affiliation{Unit{\'e} Mixte de Physique CNRS, Thales, Univ. Paris-Sud, Universit{\'e} Paris-Saclay, F91767 Palaiseau, France}             
\author{Joo-Von~\surname{Kim}}
\affiliation{Centre de Nanosciences et de Nanotechnologies, CNRS, Univ. Paris-Sud, Universit{\'e} Paris-Saclay, 91405 Orsay, France}


\begin{abstract}
A theoretical study on how synchronization and resonance-like phenomena in superparamagnetic tunnel junctions can be driven by spin-transfer torques is presented. We examine the magnetization of a superparamagnetic free layer that reverses randomly between two well-defined orientations due to thermal fluctuations, acting as a stochastic oscillator. When subject to an external ac forcing this system can present stochastic resonance and noise-enhanced synchronization. We focus on the roles of the mutually perpendicular damping-like and field-like torques, showing that the response of the system is very different at low and high-frequencies. We also demonstrate that the field-like torque can increase the efficiency of the current-driven forcing, specially at sub-threshold electric currents. These results can be useful for possible low-power, more energy efficient, applications.
\end{abstract}



\maketitle

\section{Introduction}

Thermal fluctuations in magnetic materials become progressively more important as the system dimensions are reduced toward the nanometer scale. A well-known example is superparamagnetism, where fluctuations lead to random reversals in the orientation of a magnetic nanoparticle between two bistable states~\cite{Neel:1949, Street:1949}. Such phenomena have taken on a greater importance with the advent of magnetic nanostructures, particularly in the context of magnetic data storage such as hard drive media and magnetoresistive random access memories, where much effort has been dedicated toward increasing the energy barrier separating the bistable states in order to improve thermal stability. However, such improvements can come at a cost for deterministic switching. Using current-driven spin-transfer torques~\cite{slonc1996, berger1996} (STTs), for example, the relevant thresholds for a magnetization reversal also increase with stability. This compromise between stability and switching efficiency is a fundamental and applied problem that has stimulated much research over the past few decades~\cite{rippard2011_prb}.

A different approach to be explored in spintronic devices is to see thermal fluctuations as an advantageous effect, rather than a nuisance~\cite{locatelli2014}. One possible way to make use of thermal agitation is through stochastic resonance (SR), a phenomenon in which the addition of noise to a nonlinear system confers greater sensitivity for detecting weak periodic signals~\cite{benzi1981, gammaitoni1989, gammaitoni1998}. The effect is characterized by a resonance-like response in the signal to noise ratio that appears at an optimal non-zero noise level. In biological systems, SR appears in a number of different contexts being exploited in a variety of systems in nature~\cite{wiesenfeld1995}, such as the enhancement of information transfer in the mechanoreceptors of crayfish~\cite{douglass1993} and in the enhanced neural processing in a sensory system of crickets~\cite{levin1996}. Stochastic resonance has also been observed in magnetic systems~\cite{grigorienko1995}, with recent examples involving current-driven dynamics in magnetoresistive multilayers~\cite{cheng2010, finocchio2011, daquino2011, daquino2012}. 
For nanoscale spintronic devices such as magnetic tunnel junctions (MTJs), it has been suggested that stochastic resonance and noise-enhanced synchronization might be useful tools for low-power information processing schemes that draw much of the required energy from the thermal bath of the environment~\cite{locatelli2014}.

Magnetic tunnel junctions typically comprise a free magnetic layer and a reference magnetic layer, where the latter can be a composite structure such as an exchange-biased synthetic antiferromagnet. From an information storage perspective, it would be desirable to maximize the thermal stability of the free layer magnetization, such that the probability of thermal fluctuations driving magnetization reversal remains low. On the other hand, the desired feature for a stochastic oscillator is the thermally driven transitions between the free layer magnetic states. As such, MTJs are good candidates for noise-enhanced applications, particularly at sub-100 nm scales where certain material structures can result in a superparamagnetic state at room temperature~\cite{locatelli2014}. A particular feature of current-driven torques in magnetic tunnel junctions is the presence of two components: the usual damping-like term~\cite{slonc1996, berger1996}, which represents a nonconservative torque, and a field-like torque~\cite{sankey2008, devolder2009jap, matsumoto2011} (FLT), which appears like an effective field in the equations of motion. The motivation here is to understand how the interplay between the conservative and nonconservative driving terms will influence the stochastic dynamics of such a system.

In this paper, we examine theoretically how synchronization and resonance-like phenomena in superparamagnetic tunnel junctions can be driven by spin-transfer torques and how the different torques will affect the response of the system. We consider a superparamagnetic free layer state, which reverses randomly between two well-defined orientations as a result of thermal fluctuations, giving rise to telegraph noise. Because an average frequency can be associated to such reversals, we can consider this system as a stochastic oscillator.  

The article is organized as follows. In Section II, we describe the model used and the simulation method employed to study the stochastic dynamics. In Section III, we present results of stochastic resonance and synchronization to different amplitudes of periodic input electric currents while changing the temperature. A detailed analysis on the role of the different spin-torque components is given in Section IV, where we show the existence of two different regimes: the low frequency (LF) one, where both spin-transfer torques contribute, and the high frequency (HF) regime, where the FLT appears to be the most important term driving the dynamics. Concluding remarks are presented in Section V. 

\section{Model and simulation methods}

\subsection{Model and geometry}

\begin{figure}
\centering
\includegraphics[width=8.5cm]{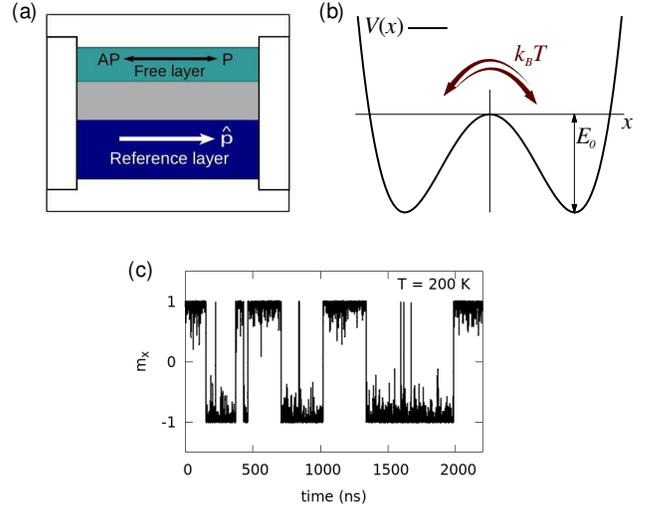}
\caption{(Color online) (a) Schematic MTJ showing the free and reference layers. The free layer magnetization has two stable positions: P, with $m_x = 1$, and AP, with $m_x = -1$. (b) A generic double well potential $V(x)$ with thermal noise. As the temperature grows, $E_0/k_B T$ gets smaller and the probability of the system going from one stable state to the other gets higher. (c) Telegraph noise of the free layer of the MTJ at $T = 200$ K shows $m_x$ oscillating between its two equilibrium points.
}
\label{fig01}
\end{figure}

The free layer of the magnetic tunnel junction is assumed to be elliptical with lateral dimensions of 150 $\times$ 50 nm and a thickness of 2 nm. We take $x$ and $y$ to represent the long and short axes of the ellipse, respectively, and $z$ to represent the direction perpendicular to the film plane. We assume that the free layer magnetization $\mathbf{M}$ is uniform and treat its dynamics in the macrospin approximation. As such, the magnetization orientation can be described by the unit vector $\mathbf{m} = (m_x, m_y, m_z) \equiv \mathbf{M}/M_s$. In the absence of any applied magnetic fields, the magnetic energy of the free layer can be expressed as 
\begin{equation} 
E({\mathbf m}) = \frac{1}{2} \mu_0 M_s^2 V \left( m^2_z - q m^2_x\right ),
\label{magEnergy} 
\end{equation} 
where $\mathbf{\hat{z}}$ is the hard axis, representing the demagnetizing fields due to the thin film geometry, and $\mathbf{\hat{x}}$ is the easy axis associated with the shape anisotropy of the elliptical dot whose strength is given by $q$, where $q \ll 1$. The system possesses two stable equilibrium points at $m_x = \pm 1$ that correspond to parallel (P) and antiparallel (AP) orientations of the free layer in respect to the reference layer [Fig.~\ref{fig01}(a)]. For the simulations considered here, we assumed a saturation magnetization of $M_s = 4 \times 10^5$ A/m, an anisotropy constant $q = 2/150$, and a volume $V = (4\pi /3) \times 1 \times 25 \times 75$ nm$^3$. These parameters lead to a sufficiently low energy barrier, $E_0$, for the MTJ such that telegraph noise is observed even at temperatures lower than room temperature, as depicted in Fig.~\ref{fig01}(c).

The telegraph noise is indicative of superparamagnetic behavior and represents a series of random reversals of the magnetization orientation between the two equilibrium points and oscillations around these points, driven entirely by thermal fluctuations. We can consider the problem as a thermally-activated particle in a double well potential with minima located at $m_x = \pm 1$ and assume that the statistical distribution of the $\mathbf{m}$ values is proportional to the Boltzmann factor $e^{-\Delta}$, with $\Delta \equiv E_0/k_B T$ and $k_B$ being the Boltzmann constant, meaning that the probability of finding $\mathbf{m}$ far from its equilibrium points is low, but that random transitions between P and AP orientations increase with $T$. This thermally-activated hopping between the two stable states can be described by an Arrhenius rate $r_T = r_0 e^{-\Delta}$, where $r_0$ is an attempt frequency related to the intra-well dynamics. Since these thermally-driven reversals of $\mathbf{m}$ persist indefinitely without any external forcing the MTJ effectively behave as a stochastic oscillator~\cite{locatelli2014}. 

The thermal macrospin dynamics is described by the stochastic Landau-Lifshitz equation~\cite{landau1935, gilbert2004} with additional spin torque terms~\cite{Slonczewski1989,slonc1996}, 
\begin{multline} 
\frac{1}{\kappa} \frac{d {\mathbf m(\tau)}}{d\tau} = - \mathbf{m}(\tau) \times \left(\mathbf{h}_{\rm eff}(\tau) + b_j(\tau)\hat{\mathbf{p}} \right) \\%
- \mathbf{m}(\tau) \times \left[\mathbf{m}(\tau) \times \left(\alpha \mathbf{ h}_{\rm eff}(\tau) + a_j(\tau)\hat{\mathbf{p}} \right) \right],  
\label{landau_gilbert_slonc} 
\end{multline} 
where $\tau \equiv \gamma M_s t$ is a dimensionless time, being $\gamma = 2.21 \times 10^5$ m/(A.s) the gyromagnetic ratio, $\alpha = 0.01$ is the Gilbert damping constant, $\kappa \equiv 1/(1+\alpha^2)$, and $\hat{\mathbf{p}}$ is a unit vector representing the orientation of the reference layer responsible for the spin polarization. The dimensionless effective field, $\mathbf{h}_{\rm eff}(\tau) = \mathbf{h}^{E}(\tau) + \mathbf{h}^{T}(\tau)$, consists of two components. The first represents the deterministic part that is associated with the magnetic energy given in Eq. (\ref{magEnergy}), 
\begin{equation} 
\mathbf{h}^E \equiv - \frac{1}{\mu_0 M_s^2 V}\nabla_{\mathbf{m}}E =  q m_x \hat{\mathbf{x}} - m_z \hat{\mathbf{z}}.
\label{hconserv} 
\end{equation} 
The second term, $\mathbf{h}^{T}(\tau)$, is the one that accounts for the thermal fluctuations. This stochastic field represents a Gaussian white noise with no correlation between its different Cartesian components $(i,j)=(x,y,z)$,
\begin{equation}
\langle h^{T}_i(\tau) \rangle = 0; \; \; \langle h^{T}_i(\tau) \, h^T_{j}(\tau') \rangle = D \, \delta_{ij} \, \delta(\tau - \tau'),
\end{equation}
where the diffusion constant, $D$, is a measure of the temperature and is determined by the fluctuation-dissipation theorem~\cite{brown1963},
\begin{equation} 
\label{eq-D} 
D = \frac{2 \alpha k_B  T}{  \mu_0 M_s^2 V }  .
\end{equation} 
In addition to the effective fields, the torques due to a time-varying spin-polarized current perpendicular to the films planes are also included~\cite{slonc1996,li_zhang2003}. The spin torques involve a field-like term characterized by the amplitude $b_j(\tau)$ and the damping-like term characterized by $a_j(\tau)$, where the latter is given by
\begin{equation}
\label{def-aj}
a_j(\tau) = \frac{\mu_B \eta I(\tau)}{\gamma |e| M_s^2 V}.
\end{equation}
Here, $\mu_B$ is the Bohr magneton, $\eta$ is the spin polarization of the current, $|e|$ is the elementary charge, $\gamma = 2.2 \times 10^5$ m/(A s) is the gyromagnetic constant, and $I(\tau)$ is the time-dependent electric current. For the remainder of the paper, the strength of the field-like torque is always taken to be a fraction of $a_j$. We assume that the spin polarization is defined by the orientation of the fixed reference layer magnetization of the magnetic tunnel junction, which is taken to be $\mathbf{\hat{p}} \equiv \mathbf{\hat{x}}$ in what follows. In this case, $a_j(\tau) > 0$ favors the parallel state, while $a_j(\tau) < 0$ favors the anti-parallel state. 

The time variation of the spin torques is defined by an oscillatory function $\xi (\tau)$, such that  $a_j(\tau) = a_j \xi (\tau)$ and $b_j(\tau) = b_j \xi (\tau)$. These terms are related to the applied current that can also be represented as the product of a fixed amplitude, $I_{AC}$, and the time dependent component as $I(\tau) = I_{AC} \xi(\tau)$. For this study, we considered a square wave function $\xi (\tau) = \pm 1$ that has an input frequency $f_{in}$. 

\subsection{Simulation methods}

We studied the stochastic macrospin dynamics by numerically integrating the equations of motion [Eq.~(\ref{landau_gilbert_slonc})] for different parameters such as the input ac frequency, temperature, the amplitude of the applied electric current (ac forcing), and the ratio $b_j/a_j$. Because Eq.~(\ref{landau_gilbert_slonc}) represents a nonlinear Langevin equation with multiplicative noise, we use the Stratonovich interpretation~\cite{garciapalacios1998,martinez2004} for the time integration of our stochastic differential equations. We verified that $|m| = 1$ is preserved to guarantee the correctness of the numerical scheme. The thermal field, $h^{T}$, is drawn from a standard normal distribution $\mathcal{N}(0,1)$,
\begin{equation} 
\label{eq-hT} 
h^{T}_i = \sqrt{\frac{D}{\Delta \tau}} \mathcal{N}(0,1),
\end{equation} 
where $\Delta \tau = 0.01$ is the integration time-step used for the numerical simulations. We employed the Mersenne Twister~\cite{matsumoto1998} algorithm for random number generation, which is implemented in the GNU Scientific Library.

The magnetization projection $m_x$ determines the tunnel magnetoresistance, which allows the use of $m_x$ as the system output to compute the average transition rate between the two stable states, which we can see as an output frequency. Because of the stochastic nature of the problem the input signal can have a well defined frequency but the output needs to be understood as an average frequency, $\langle f_{out} \rangle$, that we define as half the number of $m_x$ reversals $N_{rev}$ over the total computed time $t_T$: 
\begin{equation} 
\label{def-fout}
\langle f_{out} \rangle \equiv \frac{N_{rev}}{2t_T}.
\end{equation}
To ensure that $\langle f_{out} \rangle$ is a statistically significant average we used very long integration times and repeated several times the simulations for each data point, specially for low temperature and low frequency cases. For clarity, we keep $f_{in}$ and $\langle f_{out} \rangle$ as the input and average output frequencies measured in S.I. units but also make use of a ratio of these values to the thermal transition rate at $293$ K, $\Gamma$, for easier comparison among different dynamics. The thermal transition rate corresponds to the average number of back and forth transitions between the P and AP states in the free running system, with no applied current, counted in the same way as defined by Eq.~(\ref{def-fout}). In our case $\Gamma \approx 28$ MHz. 

It is important to check whether the input and output signals are correlated. We use the following quantity as a measure of this correlation, 
\begin{equation} 
\label{def-correlation}
\sigma = \sum_n \frac{m^n_x \cdot \xi^n (\tau)}{N},
\end{equation}
which represents the product between $x$-component of the magnetization at every numerical iteration $n$ ($m_x^n$) and the input signal $\xi^n (\tau) = \pm 1$. This is averaged over the total number of iterations $N$. Essentially the correlation will be a measure of the fraction of time that the output $m_x$ responds to the input $\xi (\tau)$ and can be used to evaluate how changing the system parameters affects this response and as a measure of stochastic resonance. 

In the following sections we present the results of the numerical simulations for several cases. The dynamical response of the system can vary depending on the presence or not of the field-like torque, $b_j$, the temperature, $T$, the input frequency, $f_{in}$, and the amplitude of the input signal, $I_{AC}$. The cases where we are changing the $b_j/a_j$ ratio are explicitly indicated on the figures. For the cases when nothing is mentioned we use a default value of $b_j = 0.3a_j$. The critical current for a magnetization reversal at null temperature 
changes if the ratio $b_j/a_j$ is changed, so we make use of two critical currents, $I_{c1}$ and $I_{c2}$, for the cases where $b_j = 0.0$ and $b_j = 0.3a_j$ respectively. At any $T \neq 0$ the current necessary to obtain a magnetization reversal will be less than the zero temperature value, which allows the use of subthreshold currents. 


\section{Stochastic Resonance and Synchronization}

\begin{figure}
\centering
\includegraphics[width=8.5cm]{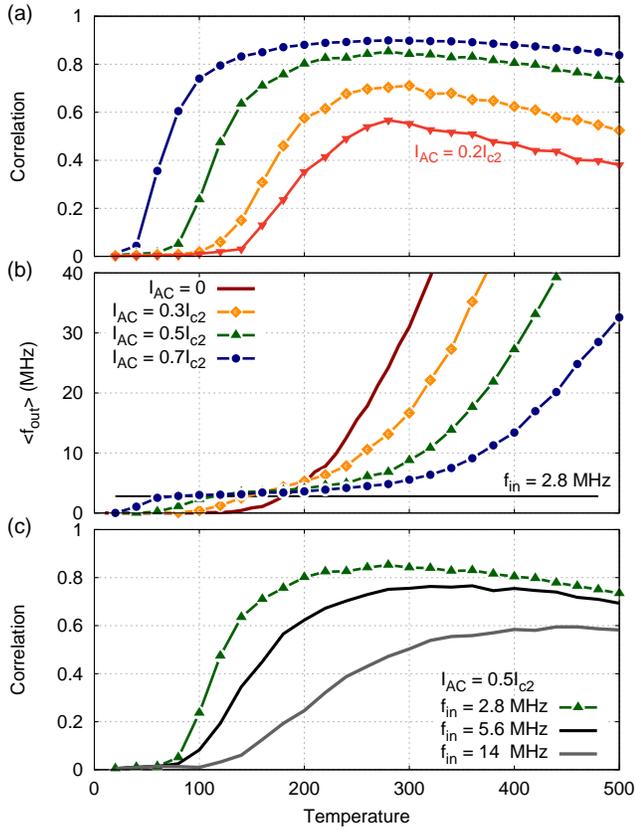}
\caption{(Color online) (a) Correlation for different amplitudes of the applied electric current ($I_{AC} = 0.2I_{c2}$, $0.3I_{c2}$, $0.5I_{c2}$ and $0.7I_{c2}$) with a constant input frequency, $f_{in} = 2.8$ MHz, as function of the temperature. As $I_{AC}$ grows the stochastic resonance curves broadens and the peak becomes less distinct. (b) Plot of $\langle f_{out} \rangle$ as function of temperature. Increasing the amplitude of the external forcing makes $\langle f_{out} \rangle$ approach the input frequency inside an interval of temperature values. (c) Correlation for $I_{AC} = 0.5I_{c2}$ and $f_{in} = 2.8$ MHz ($0.1\Gamma$), $5.6$ MHz ($0.2\Gamma$), and $14$ MHz ($0.5\Gamma$). The maximum is smaller and happens at higher temperatures as the input frequency increases.
}    
\label{fig02}
\end{figure}

We start analyzing the occurrence of stochastic resonance when the magnetic tunnel junction is subject to a subthreshold ac signal by computing the correlation $\sigma$ as a function of temperature [Fig.~\ref{fig02}(a)]. For a fixed input frequency $f_{in} = 2.8$ MHz, or $0.1\Gamma$, the correlation grows for increasing applied current, with peak values going from $\sigma \approx 0.6$ at $I_{AC} = 0.2I_{c2}$, to $\sigma \approx 0.9$ at $I_{AC} = 0.7I_{c2}$. The correlation peaks almost at the same temperature ($T \approx 280$ K) for all the amplitudes of applied current, $I_{AC} = 0.2I_{c2}$, $0.3I_{c2}$, $0.5I_{c2}$ and $0.7I_{c2}$, but the shape of the curve and the maximum correlation value will change, with the peak becoming less distinct at higher values of electric current. The temperature at which the peak happens is mainly determined by the input frequency, since SR maxima are associated to a time matching criteria~\cite{gammaitoni1998} between the average waiting times of thermally activated hopping and the oscillations of the external periodic forcing. Because of this relation, the correlation peak corresponding to smaller values of $f_{in}$ will happen at a lower temperature than the peak for higher values of the input frequency. This can be seen in Fig.~\ref{fig02}(c) where we keep the amplitude of the electric current at $I_{AC} = 0.5I_{c2}$ and increase $f_{in}$ to $5.6$ MHz ($0.2\Gamma$), and to $14$ MHz ($0.5\Gamma$). The correlation maximum is reduced and moves to higher temperature values.

%
Another important feature of this system is a locking of the output frequency to the input that can happen for some temperature interval. This frequency-locking appears as the thermally activated transitions become synchronous to the forcing frequency, which results in the average output frequency, $\langle f_{out} \rangle$, approaching $f_{in}$ for a range of temperatures, or inside a locking-region. In  Fig.~\ref{fig02}(b) we can see what happens to the average output frequency as we increase the amplitude of the ac forcing from $I_{AC} = 0$ to $I_{AC} = 0.3$, $0.5$, and $0.7I_{c2}$. As the external forcing grows, $\langle f_{out} \rangle$ deviates from the thermal transition rate ($I_{AC} = 0.0I_{c2}$) and approaches the horizontal line that marks $f_{in} = 2.8$ MHz. Not surprisingly, a higher ac amplitude value creates a better locking between the frequencies, widening the temperature interval where $\langle f_{out} \rangle \approx f_{in}$. Although $\langle f_{out} \rangle$ approaches $f_{in}$, from $T \approx 60$ K to $T \approx 200$ K for $I_{AC} = 0.7I_{c2}$, they do not match exactly because of the existence of random switching between states. This leads to additional reversals that do not follow $I(\tau)$ and increase $\langle f_{out} \rangle$. For the same reason the correlation is not perfect, i.e. $\sigma < 1$, as these additional reversal events, or ``glitches'', decrease $\sigma$.

\begin{figure}[]
\centering
\includegraphics[width=8.5cm]{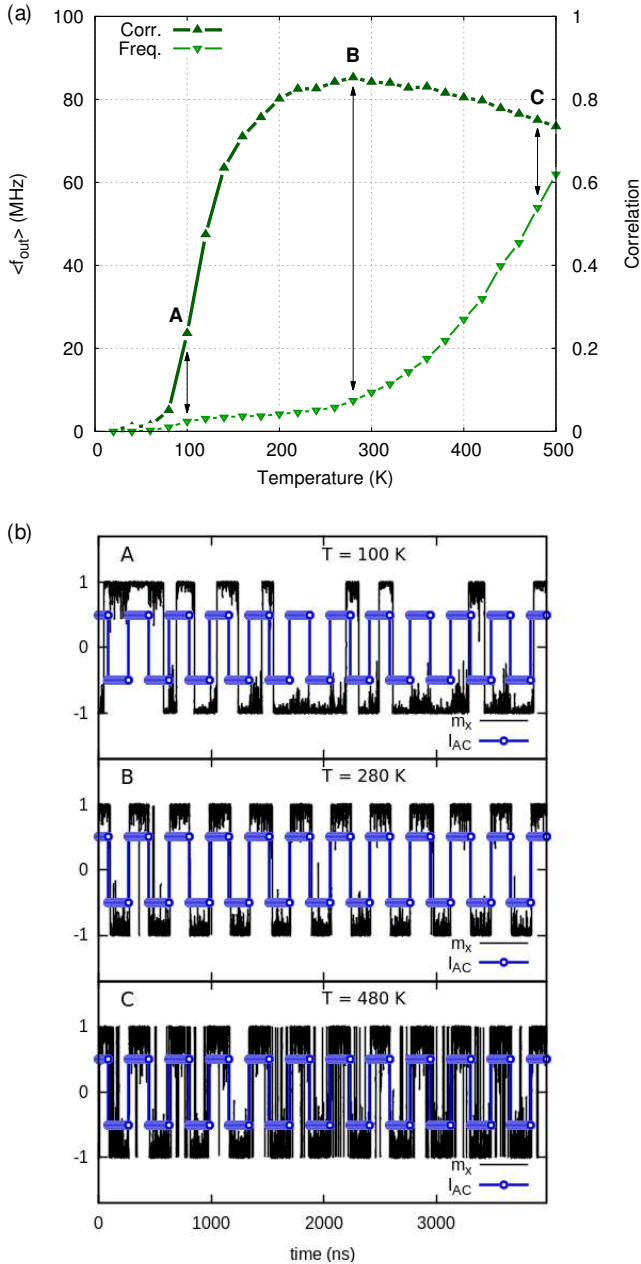} 
\caption{(Color online) (a) Correlation (Corr.) and $\langle f_{out} \rangle$ (Freq.) for $I_{AC} = 0.5I_{c2}$ and $f_{in} = 2.8$ MHz. As the temperature increases the correlation peaks at $T = 280$ K (point B) and decreases slowly, while $\langle f_{out} \rangle$ overcomes the locking region and then grows exponentially. (b) Time traces showing $m_x(t)$ and $\xi(t)/2$ at points A ($T = 100$ K), B ($T = 280$ K) and C ($T = 480$ K) for $I_{AC} = 0.5I_{c2}$.
}
\label{fig03}
\end{figure}

It is easier to understand the relation between $\langle f_{out} \rangle$ and the correlation as the temperature changes by plotting these quantities together for the same applied current, as shown in Fig.~\ref{fig03}(a). The three points A ($T = 100$ K), B ($T = 280$ K), and C ($T = 480$ K) indicate different regimes. At point A, $\sigma$ and $\langle f_{out} \rangle$ are low because at low temperature the system is rarely able to overcome the energy barrier. With increasing $T$ the correlation grows while the average output frequency stays close to the value of $f_{in} = 2.8$ MHz, until getting to point B, when the correlation peaks at $\sigma \approx 0.85$ and the frequency-locking no longer exists since $\langle f_{out} \rangle$ is now growing exponentially. Finally, at point C, the average output frequency is very far from the input frequency, but the correlation is still strong. This kind of behavior is easy to understand looking at Fig. \ref{fig03}(b), where the temporal evolution of $m_x(t)$ and $\xi(t)$ are presented. The correlation stays strong at $T = 480$ K because the glitches that increase $\langle f_{out} \rangle$ are fast, so the magnetization quickly comes back to the state favored by the input signal and the total fraction of time the system is correlated remains high.  
%
%

The mismatch between the input driving signal and the magnetization reversals can be characterized in terms of phase diffusion by examining how the relative phase between the magnetization and driving signal evolves with time~\cite{locatelli2014}. This can be cast in terms of a discrete one-dimensional random walk problem where $+1$ is assigned to the case where the driving signal and magnetization state are in phase, and $-1$ to the case when they are not. Using phase analysis it is possible to get to an analytic expression for the output frequency in terms of the amplitude of the input signal, noise, and $f_{in}$~\cite{neiman1999,freund2003}. However, the locking between output and input frequencies predicted with this formalism is better than the numerical results we obtained, with a matching $\langle f_{out} \rangle$/$f_{in}$ very close to 1 for a larger interval of temperatures, or noise. The reason for this is that the theory relies on the adiabatic limit and rests on the assumption that the input amplitude, noise amplitude, and phase difference are the only factors contributing to the output. 
In our analysis it is clear that both stochastic resonance and frequency-locking have an adiabatic character, meaning that only input frequencies lower than the natural transition rate of the system will produce strong results, but even these low frequencies values will not match exactly the adiabatic limit. 

To fully understand the behavior of the system it is necessary to take in consideration its different response to different input frequencies, which means that HF and LF regimes obey different dynamics. Higher values of $f_{in}$ will be less effective to promote correlation, having a smaller peak value that happens at a higher temperature. Changing $f_{in}$ from $2.8$ MHz to $14$ MHz reduces the peak correlation by approximately $25\%$, for example. A similar result is found when the output frequency is analyzed, considering the temperature interval where $\langle f_{out} \rangle / f_{in} \approx 1$, with the locking region becoming smaller as the input frequency grows. It is still noteworthy that even with a small ac amplitude of 20\% of the critical value [Fig.~\ref{fig02}] it is possible to reach a peak correlation of $\sigma \approx 0.6$, which is indicative of how stochastic resonance can be used in more energy efficient devices where a considerable amount of the operating power comes from the thermal bath of the environment. 

\section{Role of the different spin torque components}

In the analysis so far, we have examined cases in which the ratio between the field-like and damping-like spin torques has been kept constant, at $b_j/a_j = 0.3$.  In this section we examine the role of these components, which result in different dynamical responses of the system. The differences associated to variations of the damping-like term and of the FLT appear in the average output frequency and in the correlation and are dependent of the input frequency, so we make a separate analysis of the low frequency and the high frequency cases.

\subsection{Low Frequency}

We start taking a closer look at the low frequency regime ($f_{in} \lesssim \Gamma$), where it is possible to observe stochastic resonance, high correlation, and frequency-locking. We analyze the results produced by different $b_j/a_j$ ratios while keeping the input frequency fixed and changing the temperature and also for the opposite case of fixed temperature while changing $f_{in}$.

\begin{figure}[]
\centering
\includegraphics[width=8.5cm]{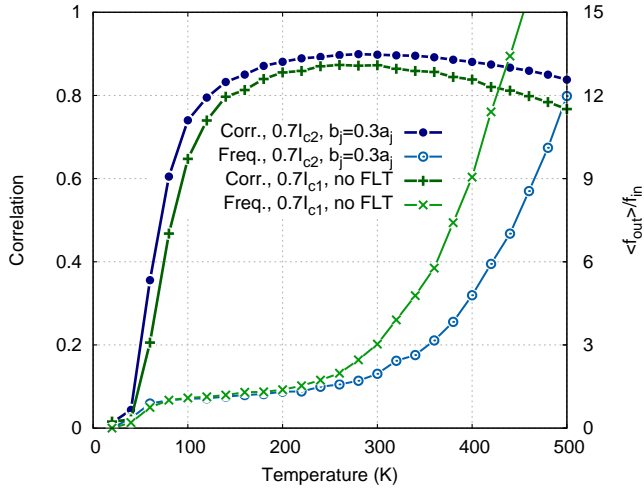}
\caption{(Color online) Correlation (Corr.) and $\langle f_{out} \rangle/f_{in}$ (Freq.) as function of temperature with ($b_j = 0.3a_j$) and without the FLT, for $f_{in} = 0.1 \Gamma$. The presence of the field-like torque increases both the correlation and the frequency-locking range, indicating a greater efficiency at similar amplitudes of the input signal.
}
\label{fig04}
\end{figure}

First we compare the correlation and average output frequency as functions of temperature for two cases: $b_j/a_j = 0.0$ and $0.3$, with the same input frequency $f_{in} = 0.1\Gamma$ (Fig.~\ref{fig04}). Since the ratio $b_j/a_j$ changes the critical current, we must use the two different currents $I_{c1}$, for $b_j/a_j = 0.0$, and $I_{c2}$, for $b_j/a_j = 0.3$, with $I_{c1} > I_{c2}$, as parameters to ensure we are seeing changes caused by the field-like torque and not simply by an increased amplitude of the applied current in respect to the critical current. In both cases we set the applied current to be 70\% of the threshold value, namely $I_{AC} = 0.7I_{c1}$ and $I_{AC} = 0.7I_{c2}$. The presence of the field-like torque increases both the correlation and the range that $\langle f_{out} \rangle$ is locked to the input, indicating a higher efficiency of the spin transfer torques at avoiding unwanted random reversals caused by thermal agitation. 

\begin{figure}[b]
\centering
\includegraphics[width=8.5cm]{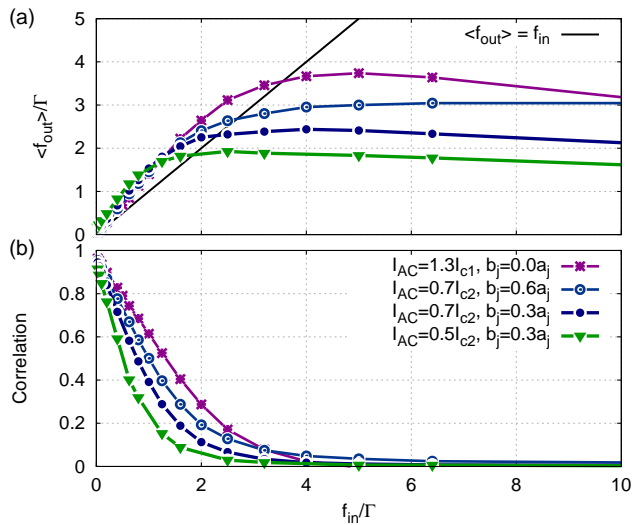}
\caption{(Color online) (a) ratio $\langle f_{out} \rangle /\Gamma$ and (b) correlation as function of $f_{in}/\Gamma$ for four different input currents at low frequency: $1.3I_{c1}$ with $b_j/a_j = 0.0$, $0.7I_{c2}$ with $b_j/a_j = 0.3$ and $0.6$, and $0.5I_{c2}$ with $b_j/a_j = 0.3$. The black line shows $\langle f_{out} \rangle = f_{in}$.  
}
\label{fig05}
\end{figure}

A known fact of stochastic resonance systems is that the resonance-like behavior that appears as the noise changes does not appear when the input frequency is the variable being changed, with the system showing a essentially monotonic response~\cite{choi1998}. We now analyze how $\langle f_{out} \rangle$ and the correlation changes as function of $f_{in}$ [Fig.~\ref{fig05} (a) and (b), respectively] for different $a_j$ and $b_j$ amplitudes at the same temperature $T = 300$ K. At very low frequencies there is a linear growth in the average output frequency with $f_{in}$ and the correlation is high, getting close to the limit $\sigma = 1$ as $f_{in} \to 0$. The correlation and the locking of the output frequency to the input will increase with increasing $I_{AC}$ and $b_j/a_j$ ratio, but quickly decrease with increasing input frequency, as can be seen from the presented curves with $I_{AC} = 1.3I_{c1}$ (no FLT), $0.7I_{c2}$ ($b_j/a_j = 0.3$ and $0.6$), and $0.5I_{c2}$ ($b_j/a_j = 0.3$). When $f_{in}$ gets close to twice the value of the average thermal transition rate $\Gamma$, the correlation is almost zero and $\langle f_{out} \rangle$ essentially no longer has a linear dependence on $f_{in}$.   
The average output and input frequencies never match exactly because of unavoidable thermally activated phase slips. Testing the cases of an above-threshold current $I_{AC} = 1.3I_{c1}$, with no field-like torque, and of an increased FLT component $b_j = 0.6a_j$, with $I_{AC} = 0.7I_{c2}$, the result is that both average output frequency and correlation will have a stronger response to larger applied currents than to an increased field-like torque at low frequency. Nevertheless, larger $b_j/a_j$ ratios require in general smaller applied currents to get similar rates of correlation and frequency-locking when compared to a lower or zero FLT.

\subsection{High Frequency}

\begin{figure}[b]
\centering
\includegraphics[width=8.5cm]{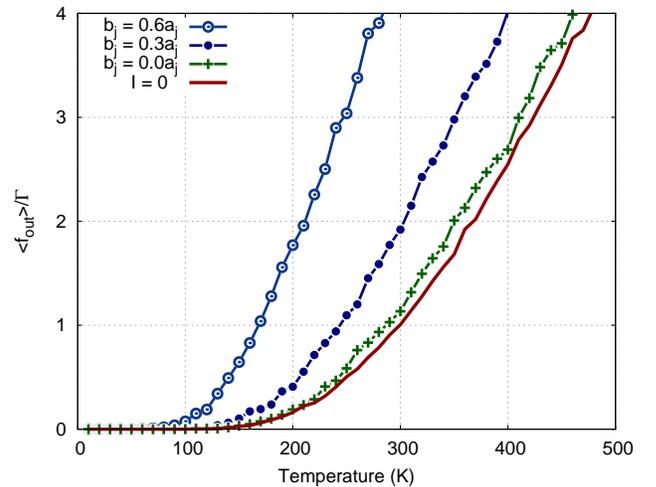}
\caption{(Color online) The average output frequency of the system increases for higher values of the FLT at high input frequency, in the case $f_{in} = 50\Gamma$. $I=0$ is the purely thermal transition rate curve, while $b_j/a_j = 0.0$ has $I_{AC} = 0.7I_{c1}$, and $b_j/a_j = 0.3$ and $b_j/a_j = 0.6$ both have $I_{AC} = 0.7I_{c2}$. No frequency-locking is observed in this regime.  
}
\label{fig06}
\end{figure}

The amplitude of the field-like torque component can change the transition rates and affect the system dynamics at low frequency but these variations become more pronounced in the high frequency regime ($f_{in} \gg \Gamma$). We start the HF analysis by evaluating $\langle f_{out} \rangle/\Gamma$ as a function of temperature [Fig.~\ref{fig06}]. Again, we plot the pure thermal transition rate curve (I=0), with no applied current, in order to compare it with three cases where the input frequency is fixed at $f_{in} = 50\Gamma$, but the electric current and the ratio $b_j/a_j$ are changed. We use the ratios $b_j/a_j = 0.0$ with $I_{AC} = 0.7I_{c1}$, $b_j/a_j = 0.3$ with $I_{AC} = 0.7I_{c2}$, and $b_j/a_j = 0.6$ with $I_{AC} = 0.7I_{c2}$ to show that increasing the ratio $b_j/a_j$, for the same input frequency, increases the average output frequency with increasing temperature, which causes a deviation of the curve from the thermal transition rate. This high frequency effect is probably caused by the field-like torque response because of the different ways that $a_j$ and $b_j$ enter in Eq.~(\ref{landau_gilbert_slonc}). 
While $b_j$ appears on the precession term (fast dynamics) of the LLG equation, the damping-like term represents a slow dynamics contribution, becoming less important at high input frequencies due to its adiabatic nature. 

\begin{figure}[tbp]
\centering
\includegraphics[width=8.5cm]{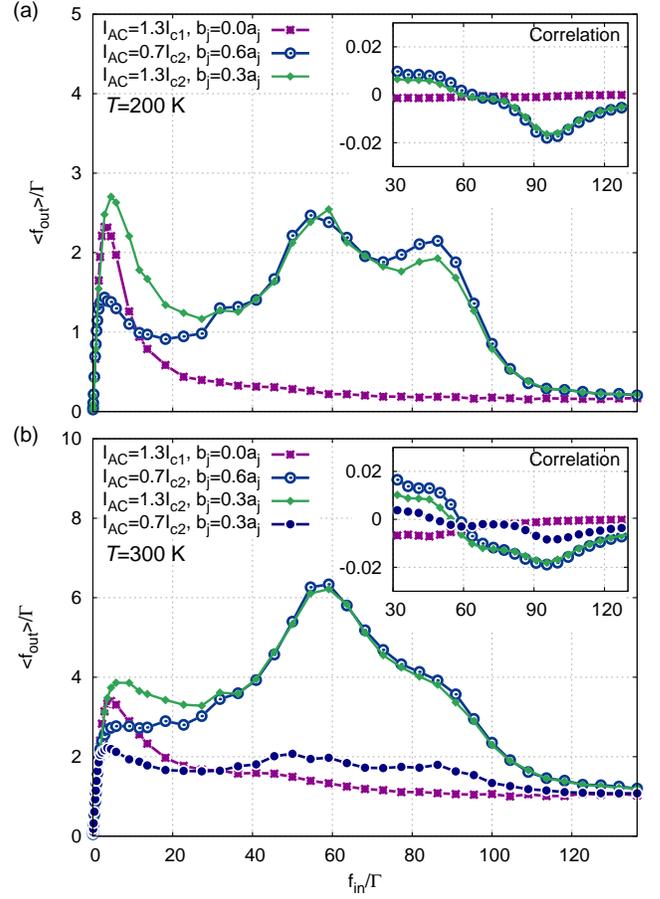}
\caption{(Color online) Plot of $\langle f_{out} \rangle/\Gamma$ as a function of $f_{in}/\Gamma$ at $T=200$ K (a) and at $T=300$ K (b) for different applied currents, and $b_j/a_j$ ratios. Three peaks are visible, with the major one happening close to $60\Gamma$, and smaller ones around $30\Gamma$ and $90\Gamma$. The peaks are associated to resonances of the natural frequencies of oscillation of $\mathbf{m}$ and their amplitudes are dependent of the parameters of the system. Insets in (a) and (b): the correlation as a function of $f_{in}/\Gamma$ shows a small negative peak close to $f_{in} = 95\Gamma$. 
}
\label{fig07}
\end{figure}

We have already shown that when the noise level is changed the system can show stochastic resonance and a locking, or synchronization, to the input signal. A similar frequency-locking also appears when varying the input frequency at LF, as shown in Fig~\ref{fig05}(a), but in this case the response of the system does not present a resonant behavior. At HF this picture changes: the frequency-locking disappears but now high amplitude resonance peaks will occur as long as the field-like torque is present in the system. This can be seen in Fig.~\ref{fig07} where we show the computed average output frequency for a wide range of $f_{in}$ values, different applied currents, $b_j/a_j$ ratios, and two different temperatures. After the initial low frequency tail, where $\langle f_{out} \rangle$ follows $f_{in}$, there is a significant increase in the output frequency, with the appearance of three peaks in the cases where $b_j$ is not null. The peaks happen around $f_{in} = 30\Gamma, 60\Gamma,$ and $90 \Gamma$, with the major resonance-like maximum close to $f_{in} = 60\Gamma$. The higher frequency peak, seen around $f_{in} = 90\Gamma$, is easier to notice in the cases where $T = 200$ K [Fig.~\ref{fig07}(a)]. These two major peaks are present only when the field-like torque is included and they grow as the ratio $b_j/a_j$ increases, indicating the influence of the FLT on the transition rates, but the first peak, at lower frequency, also seems to appear for the curve with $I_{AC} = 1.3I_{c1}$ and $b_j = 0$. 

To explain these peaks we examine the deterministic precession term of the LLG equation (\ref{landau_gilbert_slonc}), mainly responsible for the high frequency dynamics,
%
that can be written as a system of three first order non-linear coupled equations:
\begin{eqnarray}
\dot{m}_x &=& m_y m_z, \label{mxdot}\\
\dot{m}_y &=& - m_x m_z - q m_x m_z - b_j(\tau) m_z, \label{mydot}\\
\dot{m}_z &=& q m_x m_y + b_j(\tau) m_y, \label{mzdot}
\end{eqnarray}
where $b_j (\tau)$ enters directly on $\dot{m}_y$ and $\dot{m}_z$ equations but not on $\dot{m}_x$. This differs from the low frequency (damping term) of Eq.~\ref{landau_gilbert_slonc}, where $a_j$ is present in all the components. The complete system of equations cannot be solved analytically (except for some specific cases), so we will concentrate on equations (\ref{mxdot}) to (\ref{mzdot}). The two different cases of high frequency precession involve in-plane oscillations around the stable points $|m_x| = 1$ and out-of-plane large amplitude oscillation with $m_z$ close to an average value $\langle m_{z1} \rangle$ and $m_x$ and $m_y$ ranging from $-1$ to $+1$. Therefore, we can try to find the ``natural frequencies'' $\omega_0$ and $\omega_1$ for this cases, remembering to understand the results as an average approximation due to the stochastic nature of the problem. 

On the first case we can consider $m_x \approx 1$ and set $b_j = 0$. This makes $m_y$ and $m_z$ obey two harmonic oscillator equations 
\begin{equation}
\ddot{m}_{y,z} = - q(1 + q){m}_{y,z}
\label{myzharmonic}
\end{equation} 
with natural frequency given by Kittel's formula $\omega_0 = \sqrt{q(1+q)} \approx 0.116$ that corresponds to $\approx 53\Gamma = 1.48$ GHz. This is probably associated to the larger amplitude peak because $b_j(\tau)$ is able to amplify the $m_{y,z}$ oscillations, which increases the chances of a magnetization reversal and consequently also increases $\langle f_{out} \rangle$. For the second case we can proceed in the same way, except that now we consider $m_z = \langle m_{z1} \rangle$ and set $\dot{m}_z \approx 0$. This means that these reversals happen with $m_z$ keeping an average value $\langle m_{z1} \rangle$ while rotating about the $z$ axis, forced by the demagnetizing field. The two harmonic oscillator equations generated for $m_x$ and $m_y$ are
\begin{equation}
\ddot{m}_{x,y} = - \langle m_{z1} \rangle^2(1 + q){m}_{x,y}
\label{mxyharmonic}
\end{equation} 
with a natural frequency $\omega_1 = \langle m_{z1} \rangle \sqrt{(1+q)}$. The numerical results show that during this type of motion $\langle m_{z1} \rangle \approx 0.066$ giving $\gamma M_s \omega_1/(2\pi) \approx 30\Gamma$. Therefore $2\omega_1$ corresponds to approximately $60\Gamma$ and $3\omega_1$ to $\approx 90\Gamma$. Together with the value obtained for $\omega_0$ these results can account for the HF resonance peaks as a combination of influences of in-plane and out-of-plane oscillations induced by the field-like and damping-like torques that increase the transitions between P and AP magnetization states. The effect is similar to the non-adiabatic stochastic resonance reported in Ref.~\cite{cheng2010}, although the cause is different. These fast out-of-plane oscillations are not constant, but occur in short bursts that are assisted by $b_j(\tau)$ oscillations that oppose $\dot{m_z}$ and stabilize $m_z$ around $\langle m_{z1} \rangle$, allowing several reversals to occur before the magnetization relaxes toward one of the equilibrium points. Also, because of the high frequency of the input signal, variations in $a_j(\tau)$ and $b_j(\tau)$ are faster than $m_x$ reversals, creating a time delay between them. This kind of behavior causes the small peak with a negative correlation, seen in the insets of Fig.~\ref{fig07}, close to $f_{in} = 95\Gamma$. 

To verify that these results are not caused only by the amplitude of the applied current we tested the cases of above-threshold currents $I_{AC} = 1.3I_{c1}$ with $b_j = 0.0$ and $I_{AC} = 1.3I_{c2}$ with $b_j = 0.3a_j$ and analyzed the correlation as function of the input frequency for the same cases. It is important to notice that the curves obtained with $I_{AC} = 0.7I_{c2}$ and $b_j = 0.6a_j$, and with $I_{AC} = 1.3I_{c2}$ and $b_j = 0.3a_j$ are very similar at both temperatures. This means that the amplitude of the field-like torque is the meaningful variable in this case, rather than the ratio $b_j/a_j$. If we take $(I_{AC}/I_{c2})(b_j/a_j)$ as the FLT amplitude we get similar values for these cases, so both curves in Fig.~\ref{fig07} have almost the same field-like torque amplitude. This is the reason for different system parameters creating almost the same response.

Although high input frequencies can increase $\langle f_{out} \rangle$ this effect is limited by the capacity of the system to respond to the external forcing, meaning that for very high frequency values the spin torques will no longer affect the system and the transition rates between states will be caused only by thermal fluctuations. This can be seen in Fig.~\ref{fig07} when for $f_{in} > 100\Gamma$ the average output frequency approaches the purely thermal rates for $T = 300$ K and $T = 200$ K while the correlation tends to zero. The same kind of behavior can be obtained when $\langle f_{out} \rangle$ is computed for different input frequencies, and fixed applied current and ratio $b_j/a_j$ while changing the temperature. By performing this for $f_{in} = 10\Gamma$, $50\Gamma$, $60\Gamma$, and $100\Gamma$, one observes that $\langle f_{out} \rangle$ deviates from the thermal transition rate curve, increasing from $f_{in} = 10\Gamma$ to $f_{in} = 60\Gamma$, but that a further increase to $f_{in} = 100\Gamma$ makes the average output frequency drop and approach the thermal transition rate, in good agreement to what happens when the input frequency is the variable, as in Fig.~\ref{fig07}. These results demonstrate that the external forcing that the system is subjected must be some function of $a_j$, $b_j$, and $f_{in}$ and that at really high frequencies even the field-like torque becomes unable to influence the system dynamics, after an interval where it is the dominating term. 

\section{Discussion and Concluding Remarks}

We have studied the role of different noise levels on the stochastic dynamics of the free layer magnetization. In the equations of motion, this noise enters as a random field and its amplitude depends on the temperature. However, this approach assumes that the magnetic parameters, such as the saturation magnetization or anisotropy, are independent of the temperature. While it is known that such quantities can possess a nontrivial dependence on the temperature, such variations are material dependent and it is beyond the scope of this article to account for them quantitatively. Our results could be directly applied to cases in which thermal noise is introduced by other means, such as through the application of an external random field. Experimental results demonstrate that the free layer of a superparamagnetic tunnel junction exhibits frequency-locking when electric noise, playing the role of thermal agitation, is injected in the junction~\cite{mizrahi-2016-SciRep}.

Our model is based on the assumption that the spin torques do not possess any additional angular dependence beyond the vector products given in Eq.~\ref{landau_gilbert_slonc}, i.e., $a_j$ and $b_j$ are assumed to not depend on $\mathbf{m} \cdot \hat{\mathbf{p}}$, which is the case for a constant applied voltage in magnetic tunnel junctions~\cite{slonc-2007-j3m}. In case of asymmetry in the torques this could be at least partially compensated by an external bias. In general, external bias (fields or voltages) can create or compensate asymmetries on the transition rates between P and AP states or on the critical value of applied current for magnetization reversal. Although not considering asymmetry scenarios, the work provides a description for different scenarios involving the field-like and damping-like torques, so we expect that the essential features are captured here for realistic devices. 

The choices for the size of the junction and the values of material parameters were influenced by experiments~\cite{locatelli2014} and energy considerations. The macrospin approximation should not be a problem, specially because superparamagnetic tunnel junctions can be scaled down~\cite{locatelli-2013-NatMat} to (or below) 10 nm in which case the macrospin approximation should apply. At any rate, the important point here is to understand, in general terms, the interplay between thermal noise and spin-torques in stochastic phenomena involving bistable systems.

In summary, we have studied phenomena related to the stochastic dynamics in magnetic tunnel junctions using numerical simulations. We have shown that the free layer magnetization, in a superparamagnetic state, can respond effectively to subthreshold ac input currents presenting noise-enhanced synchronization and stochastic resonance whilst exhibiting additional features related to the precessional dynamics of the magnetization. We have also demonstrated the importance of field-like spin torques in addition to their damping-like counterparts, where distinct low and high-frequency response can be observed. These results will be useful for possible low-power applications in which such stochastic oscillators can be used for new information processing paradigms.


\begin{acknowledgments}
The authors acknowledge financial support from the Agence Nationale de la Recherche (France) under Contract No. ANR-14-CE26-0021. AA acknowledges support from the Conselho Nacional de Desenvolvimento Cient{\'i}fico e Tecnol{\'o}gico (CNPq, Brazil) under Contract No. 245555/2012-9 for funding his stay in France. NL acknowledges support by a public grant overseen by the Agence Nationale de la Recherche as part of the ``Investissements d'Avenir'' program (Labex NanoSaclay, Contract No. ANR-10-LABX-0035). 
\end{acknowledgments}

\bibliographystyle{apsrev}
\bibliography{acciolySR}

\end{document}